\documentclass[preprint,12pt]{elsarticle}
\usepackage{amsmath,amssymb,amsfonts}
\usepackage{graphicx}
\usepackage{epsfig}
\usepackage{slashed}
\usepackage{a4wide}
\usepackage{fancyhdr}
\usepackage{subfigure}

\allowdisplaybreaks[1]

\newcommand{\beq}{\begin{equation}}
\newcommand{\eeq}{\end{equation}}
\newcommand{\beqa}{\begin{eqnarray}}
\newcommand{\eeqa}{\end{eqnarray}}

\newcommand{\Lagr}{\mathcal{L}}

\begin{document}

\renewcommand{\theequation}{\arabic{equation}}
\newcommand{\evl}{e^{\frac{i\stackrel{\leftarrow}{\partial}_{\beta}v^{\beta}+m}{\Lambda}}}
\newcommand{\evr}{e^{\frac{-v^{\beta}i\stackrel{\rightarrow}{\partial}_{\beta}+m}{\Lambda}}}
\newcommand{\esl}{e^{\frac{i\stackrel{\leftarrow}{\partial}_{\beta}v^{\beta}+m}{\Lambda}}}
\newcommand{\esr}{e^{\frac{-v^{\beta}i\stackrel{\rightarrow}{\partial}_{\beta}+m}{\Lambda}}}

\newcommand{\err}{e^{\frac{-v^{\beta}i\stackrel{\rightarrow}{\partial}_{\beta}+m}{\Lambda}}}

\newcommand{\Gv}{\frac{g_{\omega}}{\Lambda}}
\newcommand{\Gs}{\frac{g_{\sigma}}{\Lambda}}

\newcommand{\evnml}{e^{\frac{E}{\Lambda}}}
\newcommand{\esnml}{e^{\frac{E}{\Lambda}}}

\newcommand{\evnm}{e^{-\frac{E-m}{\Lambda}}}
\newcommand{\esnm}{e^{-\frac{E+m}{\Lambda}}}

\newcommand{\Gva}{\frac{g_{\omega}}{\Lambda}}
\newcommand{\Gsa}{\frac{g_{\sigma}}{\Lambda}}

\newcommand{\Gvb}{\frac{g_{\omega}}{ (\Lambda)^{2}}}
\newcommand{\Gsb}{\frac{g_{\sigma}}{ (\Lambda)^{2}}}
\newcommand{\panda}{$\overline{\mbox P}$ANDA~}

\begin{frontmatter}

\title{How Deep is the Antinucleon Optical Potential at FAIR energies}

\author{T.~Gaitanos, M.~Kaskulov, H.~Lenske}
\address{
Institut f\"ur Theoretische Physik, Universit\"at Giessen, 
Heinrich-Buff-Ring 16, 35392 Giessen, Germany
}
\begin{abstract}
The key question in the interaction of antinucleons in the nuclear medium 
concerns the deepness of the antinucleon-nucleus optical potential. In this work we study 
this task in the framework of the non-linear derivative (NLD) model which describes 
consistently bulk properties of nuclear matter and Dirac phenomenology of 
nucleon-nucleus interactions. We apply the NLD model to antinucleon interactions in 
nuclear matter and find 
a strong decrease of the vector and scalar self-energies in energy and density and 
thus a strong suppression of the optical potential at zero 
momentum and, in particular, at FAIR energies. This is in agreement with available 
empirical information and, therefore, resolves the issue concerning 
the incompatibility of G-parity arguments in relativistic mean-field (RMF) models. 
We conclude the relevance of our results for the future activities at FAIR. 
\end{abstract}


\end{frontmatter}

\section{\label{sec1}Introduction}

The in-medium nucleon-nucleon interaction has been an object of 
intensive theoretical and experimental research of modern nuclear physics over 
the last few decades, see for a review~\cite{EoS}. The main finding was a 
softening of the nuclear equation of state at densities reached in 
intermediate energy nucleus-nucleus collisions, which was consistent with 
a variety of phenomenological~\cite{pheno} and microscopic~\cite{micro} 
models. In addition the empirical saturation of the 
proton-nucleus optical potential turned out to be consistent with heavy-ion 
theoretical studies \cite{cass1}. 

While the bare antinucleon-nucleon ($\overline{N}N$) interaction has been 
actively studied, see Refs.~\cite{lear} and references therein, 
empirical information on the in-medium interactions of antinucleons is still very poor. 
Antiproton production has been investigated theoretically in reactions induced by protons 
\cite{cass2a} and heavy ions in the SIS-energy region 
\cite{cass2}, where some data on antiprotons were available. Complementary studies
of antiproton annihilation in nuclei~\cite{oset} and
antiprotonic atoms~\cite{gal} provided further insight on the optical potential 
at very low energies, however, with rather big uncertainties in the nuclear interior 
due to the strong annihilation cross section at the surface of the nucleus.

In the near future the FAIR facility intends to study the still controversial 
and empirically less known high energy domain of the (anti)nuclear interactions in more
details than before. 
For instance, the nuclear equation of state for strangeness degrees of freedom and also the 
in-medium antinucleon-nucleon interaction are some of the key
projects~\cite{panda_big}. They are relevant for the formation 
of exotic (anti)matter systems such as double-strange 
hypernuclei and $\overline{\Lambda}$-hypernuclei in antiproton-induced reactions 
in the \panda experiment at FAIR~\cite{panda}.

The microscopic Brueckner-Hartree-Fock calculations of the in-medium $\overline{N}N$-scattering have 
been carried out in~\cite{abhf}. 
On the other hand, a complementary theoretical background for phenomenological models builds the 
relativistic hadrodynamics (RHD). It is based on the relativistic mean-field 
(RMF) theory, which is a well established tool for infinite and finite nuclear 
systems~\cite{wal74}. However, as already shown many years ago~\cite{cass2}, 
there are still unresolved problems in RMF models, when applying them 
to antiproton-nucleus scattering and to heavy ion collisions. By just imposing 
G-parity arguments, like in microscopic models~\cite{heidenbauer,abhf}, 
the RMF do not describe the experimental data~\cite{cass2,larionov,mishustin}. 
This incompatibility of mean-field models with respect to G-parity symmetry has been also shown in recent 
transport studies~\cite{larionov}, where one had to largely decrease the 
antinucleon-meson couplings by hand in order to reproduce the empirical data.

In this work we address this issue why the
conventional RMF models do not describe antiproton-nucleus Dirac phenomenology. 
To be more specific, our studies are based on the non-linear derivative (NLD) model 
\cite{nld} to RMF. The NLD model describes simultaneously the density dependence 
of the nuclear equation of state and  the energy dependence of the 
proton-nucleus optical potential. Latter feature is missing in standard RMF models. 
Then applying G-parity transformation it is shown that 
the real part of the proton \textit{and simultaneously} the real part of the 
antiproton optical potentials are reproduced fairly well in comparison with 
phenomenological studies. We finally make predictions for the 
deepness of the real part of the antiproton optical potential and estimate its 
imaginary part at low energies and energies relevant for the forthcoming experiments at FAIR.

\section{\label{sec2}NLD formalism}

The NLD approach~\cite{nld} to nuclear matter is based essentially on the Lagrangian 
density of RHD~\cite{wal74}. It describes the interaction of nucleons through the exchange 
of auxiliary meson fields (Lorentz-scalar, $\sigma$, and Lorentz-vector 
meson fields $\omega^{\mu}$)~\cite{dbhf}
\begin{equation}
\Lagr = \Lagr_{Dirac} + \Lagr_{mes} + \Lagr_{int}
\;. \label{NDC-free}
\end{equation}
The Lagrangian in Eq.~(\ref{NDC-free}) consists of the free Lagrangians for the Dirac field 
$\Psi$ and for the meson fields $\sigma$ and $\omega^{\mu}$. The isovector 
meson $\rho$ is not considered here, for simplicity. 

In conventional RHD the interaction Lagrangian ${\cal L}_{int}$ contains meson fields 
which couple to the Dirac field  via the corresponding Lorentz-density operators 
$g_{\sigma}\overline{\Psi}\Psi\sigma$  and 
$-g_{\omega}\overline{\Psi}\gamma^{\mu}\Psi\omega_{\mu}$
for the scalar and vector parts, respectively. Such interactions
describe rather successfully the saturation properties of 
nuclear matter, but they miss the energy dependence of the mean field. 
A possible solution to this problem has been proposed in~\cite{cass2a}
where the momentum-dependent phenomenological form factors were introduced. 
In~\cite{nld} this idea has been generalized in a manifestly covariant way. 
In particular, the symmetrized interaction in the NLD model is given by
\begin{align}
{\cal L}_{int} & = 
\frac{g_{\sigma}}{2}
	\left[
	\overline{\Psi}
	\, \stackrel{\leftarrow}{{\cal D}}
	\Psi\sigma
	+\sigma\overline{\Psi}
	\, \stackrel{\rightarrow}{{\cal D}}
	\Psi
	\right]
-  \frac{g_{\omega}}{2}
	\left[
	\overline{\Psi}
	 \, \stackrel{\leftarrow}{{\cal D}}
	\gamma^{\mu}\Psi\omega_{\mu}
	+\omega_{\mu}\overline{\Psi}\gamma^{\mu}
	\, \stackrel{\rightarrow}{{\cal D}}
	\Psi
	\right]
\;. 
\label{NDC}
\end{align}
The interaction between the Dirac
and the meson fields has a similar 
functional form as in standard RHD \cite{wal74}. However, now new operators ${\cal D}$ 
acting on the nucleon fields appear, which are the non-linear functionals 
of partial derivatives
\begin{equation}
\stackrel{\rightarrow}{{\cal D}}
:= \exp{\left(\frac{-v^{\beta}i\stackrel{\rightarrow}{\partial}_{\beta}+m}{\Lambda}\right)}
~,~
\stackrel{\leftarrow}{{\cal D}}
:= \exp{\left(\frac{i\stackrel{\leftarrow}{\partial}_{\beta}v^{\beta}+m}{\Lambda}\right)}
\;. \label{ope}
\end{equation}
In Eq.~(\ref{ope}) $v^{\beta}$ denotes a dimensionless auxiliary $4$-vector and
$\Lambda$ stands for the cut-off parameter. The latter  
has been adjusted to the saturation properties of nuclear
matter~\cite{nld}. In the limiting case 
of $\Lambda\rightarrow\infty$ the standard Walecka model is retained. 

The NLD Lagrangian $\mathcal{L}$ is a functional of not only $\Psi$, $\overline{\Psi}$ and 
their first derivatives, but it depends on all higher order covariant derivatives 
of $\Psi$ and $\overline{\Psi}$. For such a generalized functional 
the Euler-Lagrange equations take the form \cite{nld}
\begin{align}
\frac{\partial{\cal L}}{\partial\phi}
-
 \partial_{\alpha_{1}}\frac{\partial{\cal L}}{\partial(\partial_{\alpha_{1}}\phi)}
&+
 \partial_{\alpha_{1}}\partial_{\alpha_{2}}\frac{\partial{\cal L}}{\partial(\partial_{\alpha_{1}}\partial_{\alpha_{2}}\phi)}
 + \cdots  + \\
&(-)^{n}\partial_{\alpha_{1}}\partial_{\alpha_{2}}\cdots\partial_{\alpha_{n}}
\frac{\partial{\cal L}}
{\partial(\partial_{\alpha_{1}}\partial_{\alpha_{2}}\cdots\partial_{\alpha_{n}}\phi)}=
0 \nonumber
\;. \label{Euler}
\end{align}
Contrary to the standard expressions for the Euler-Lagrange equation, now infinite series 
of terms ($n\rightarrow\infty$) proportional to higher order derivatives of the Dirac field
$(\phi=\Psi,\overline{\Psi})$ appear. They can 
be evaluated by a Taylor expansion of the non-linear derivative operators~(\ref{ope}). 
As shown in \cite{nld}, in nuclear matter an infinite series of terms 
can be resumed exactly and the following Dirac equation is obtained 
\begin{equation}
\left[
	\gamma_{\mu}(i\partial^{\mu}-\Sigma^{\mu}) - 
	(m-\Sigma_{s})
\right]\Psi = 0\;,
\label{Dirac}
\end{equation}
with Lorentz-vector and Lorentz-scalar self-energies defined as follows
\begin{equation}
\Sigma^{\mu}  = g_{\omega}\omega^{\mu}\evr 
~,~
\Sigma_{s} = g_{\sigma}\sigma\esr
\;. \label{Sigma}
\end{equation}
The Proca and Klein-Gordon equations for the meson fields can be also derived
\begin{align}
\partial_{\mu}F^{\mu\nu} + m_{\omega}^{2}\omega^{\nu} &=
\frac{1}{2}g_{\omega}
\left[
	\overline{\Psi}\evl \gamma^{\nu}\Psi + \overline{\Psi}\gamma^{\nu}\evr \Psi
\right],
\label{omega_meson} \\
\partial_{\mu}\partial^{\mu}\sigma + m_{\sigma}^{2}\sigma &=
\frac{1}{2}g_{\sigma}
\left[
	\overline{\Psi}\evl \Psi + \overline{\Psi}\evr \Psi
\right]
\;, \label{sigma_meson}
\end{align}
with the field tensor $F^{\mu\nu}=\partial^{\mu}\omega^{\nu}-\partial^{\nu}\omega^{\mu}$. 
The meson field equations~(\ref{omega_meson}) and~(\ref{sigma_meson}) show a similar form as in 
the linear Walecka model of RHD, except of the highly non-linear behavior of the source 
terms, which generate selfconsistent couplings between the meson-field equations.

Applying the usual RMF approximation to the idealized system of infinite nuclear matter, 
the Dirac equation (\ref{Dirac}) maintains its original form. However, we have to 
distinguish between nucleons ($N$) forming the nuclear matter  and 
antinucleons ($\overline{N}$) which interact with the nuclear matter. 
For the description of antiparticles we require G-parity
 invariance of the Dirac equation and then follow the standard procedure of applying a 
G-parity transformation ${\rm G}={\rm C}e^{i\pi I_{2}}$ to the negative energy
states, where $I_{2}$ is the operator associated with the 2nd component 
of the isospin "vector" and C is the charge conjugation 
operator. The invariance of the Dirac equation under charge conjugation
requires that the auxiliary vector $v^{\beta}$ must be odd under 
C-parity transformation. With our choice of $v^{\beta}=(1,\vec{0}\;)$ for
  positive energy solutions~\cite{nld} this results in $v^{\beta}=(-1,\vec{0}\;)$
  for the charge conjugated Dirac field. 
This leads to the following Dirac equations for nucleons 
\begin{eqnarray}
\left[
	\gamma_{\mu}(i\partial^{\mu}-\Sigma^{\mu}) - 
	(m-\Sigma_{s})
\right]\Psi_{N} & = & 0
\label{Dirac_p}
\end{eqnarray}
and antinucleons 
\begin{eqnarray}
\left[
	\gamma_{\mu}(i\partial^{\mu}+\Sigma^{\mu}) - 
	(m-\Sigma_{s})
\right]\Psi_{\overline{N}} & = & 0
\label{Dirac_pbar}
\end{eqnarray}
interacting with nuclear matter, 
where $\Psi_{N}=\Psi^{+}$ and $\Psi_{\overline{N}}=\Psi_{C}$ denote the positive energy and 
the charge conjugated Dirac fields, respectively.  

The nucleon and 
antinucleon self-energies entering Eqs.~(\ref{Dirac_p}) and~(\ref{Dirac_pbar}) are the same
\begin{eqnarray}
\Sigma_{v}\equiv \Sigma^{0} &=&
g_{\omega}\omega_{0}e^{-\frac{E-m}{\Lambda}}\;, \nonumber \\
\Sigma_{s} &=& g_{\sigma}\sigma e^{-\frac{E-m}{\Lambda}}
\;. \label{SelfenNM}
\end{eqnarray}
However, note the opposite signs in the Lorentz-vector interactions in Eqs.~(\ref{Dirac_p}) and
(\ref{Dirac_pbar}). 
Furthermore, the single particle energies $E$ have to be obtained from the
in-medium mass-shell conditions which are different for nucleons ($N$) and antinucleons ($\overline{N}$)
\begin{equation}
E_{N}(p) = \sqrt{p^{2}+m^{*2}}+\Sigma_{v}~,~~
E_{\overline{N}}(p) = \sqrt{p^{2}+m^{*2}}-\Sigma_{v}
\;. \label{mass-shel}
\end{equation}
The in-medium (or effective) Dirac mass in Eq.~(\ref{mass-shel}) is given 
by $m^{*}=m-\Sigma_{s}$. Note, that $m^{*}$  depends explicitly on particle momentum. 
Again, in the limiting case of $\Lambda\rightarrow\infty$, 
the exponential factor is equal to unity and the equations are reduced to the 
ones from the Walecka model. In the NLD model the 
cut-off parameter $\Lambda$ is of natural size, i.e., of typical 
hadronic mass scale in this problem. In the following, $\Lambda=770$~MeV is
chosen, as in the original work \cite{nld}. 

In nuclear matter the NLD equations of motion for $\omega$ and $\sigma$ 
simplify to standard algebraic equations
\begin{equation}
m_{\omega}^{2}\omega^{0} = g_{\omega}\rho_{v}
~,~
m_{\sigma}^{2}\sigma = g_{\sigma}\rho_{s} 
\; \label{mesonsNM}
\end{equation}
with the corresponding density sources $\rho_{v} = \langle \overline{\Psi}_{N} \gamma^{0}
e^{-\frac{E-m}{\Lambda}}\Psi_{N}\rangle$
and 
$\rho_{s} = \langle \overline{\Psi}_{N} e^{-\frac{E-m}{\Lambda}}\Psi_{N}\rangle$. 
The vector density $\rho_{v}$ is not related to the conserved nucleon density 
$\rho$. It has to be derived from a generalized Noether-theorem \cite{nld} and reads
\begin{align}
J^{0}  \equiv \rho = \langle \overline{\Psi}_{N}\gamma^{0}\Psi_{N} \rangle
\label{rhoBarOld}
 +   \frac{g_{\omega}}{\Lambda}
	\langle \overline{\Psi}_{N}\gamma^{0}\evnm \Psi_{N} \rangle \omega_{0}
 - \frac{g_{\sigma}}{\Lambda}
	\langle \overline{\Psi}_{N}\evnm \Psi_{N} \rangle \sigma
\quad .
\end{align}

The meson-nucleon couplings $g_{\omega}$ and $g_{\sigma}$ can be taken from any 
linear Walecka model, e.g., \cite{wal74}, as it has been done here. Moreover, we use 
the same coupling constants for both nucleon and antinucleon interactions.

\section{Results and Discussion}

        \begin{figure*}[t]
        \begin{center}
        \includegraphics[width=0.7\linewidth]{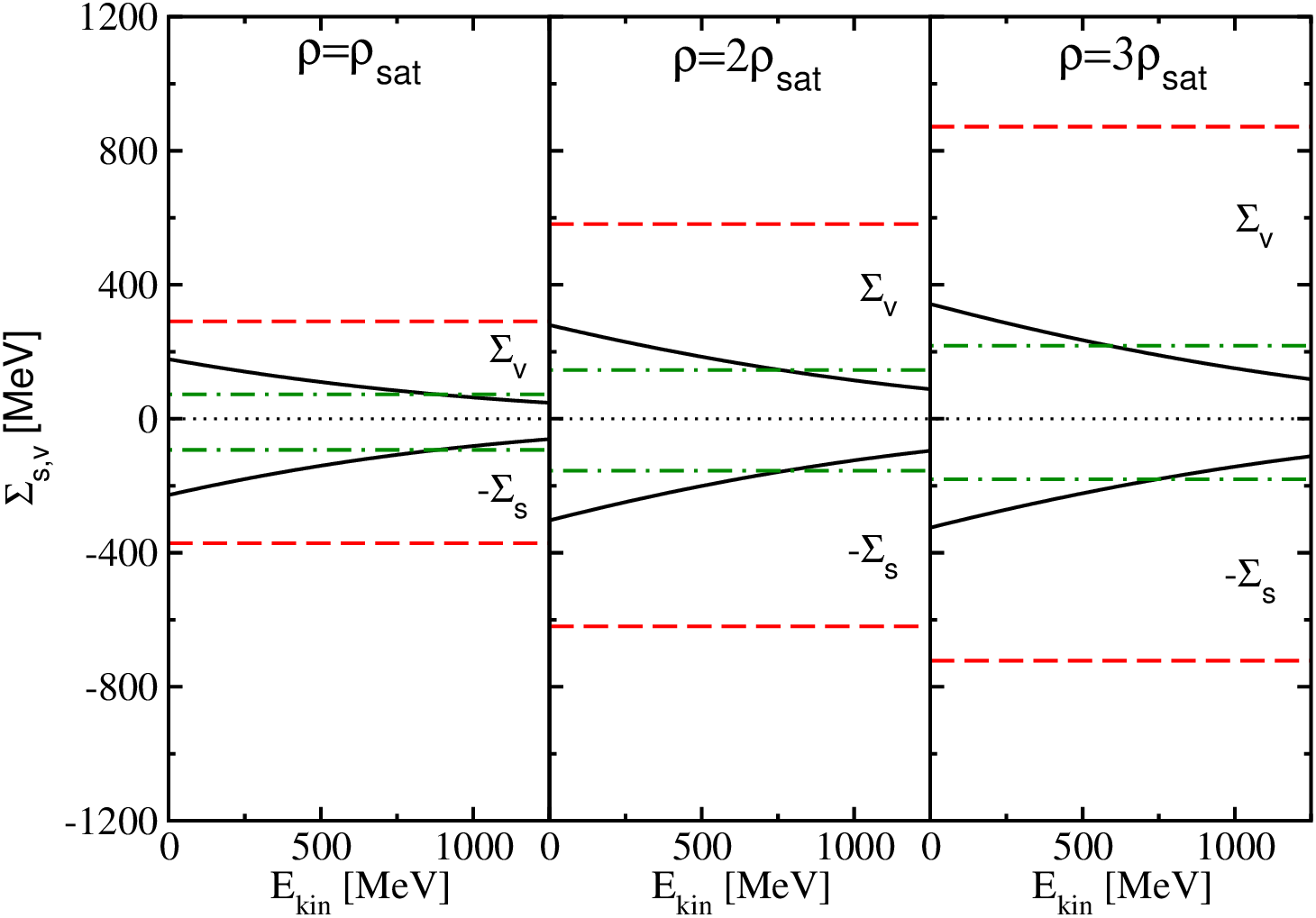}
        \end{center}
        \vspace{-0.2cm}
        \caption{Kinetic energy dependence of the scalar and vector Lorentz-components 
of the antinucleon self-energy in nuclear matter at densities of 
$\rho=\rho_{\rm sat}$ (left), $\rho=2\rho_{\rm sat}$ (middle) and
$\rho=3\rho_{\rm sat}$ (right)
using the linear Walecka model (dashed lines), the linear Walecka model with rescaled 
couplings with the factor $\xi=0.25$~\cite{larionov} (dash-dotted) 
and the NLD approach (solid lines). }
        \label{fig1}
        \end{figure*}

We have applied both the NLD and the conventional linear  Walecka models 
to nuclear matter at various baryon densities and also at various 
nucleon and antinucleon energies relative to matter at rest. At first, 
we discuss the self-energies, which are real quantities in RMF. Then we focus our study 
on the energy and density dependencies of the optical potential, first for
in-medium proton interactions, 
and then for the antiproton case.

Fig.~\ref{fig1} shows the Lorentz-scalar and Lorentz-vector components of the antinucleon 
self-energy in nuclear matter, $\Sigma_{s}$ and $\Sigma_{v}$, as a function of the kinetic 
energy at three baryon densities. The antinucleon kinetic energy is calculated relative to the potential 
depth of the nuclear matter at rest, i.e., $E_{kin}=E_{\overline{N}}-m=\sqrt{p^{2}+m^{*2}}-\Sigma_{v}-m$. 
The NLD calculations show an explicit energy dependence for both components of the 
antinucleon self-energy. In particular, the self-energies decrease 
with increasing energy, for all baryon densities. On the other hand, with rising baryon 
density they increase only moderately at each energy. The saturation in energy and density 
results from the non-linear interaction, as discussed in detail in Ref.~\cite{nld}. 
In the linear Walecka model the Lorentz-vector self-energy grows strongly with increasing 
density, while the Lorentz-scalar component saturates. Both components in the standard RMF 
are energy independent. 

        \begin{figure*}[t]
        \begin{center}
        \includegraphics[width=0.7\linewidth]{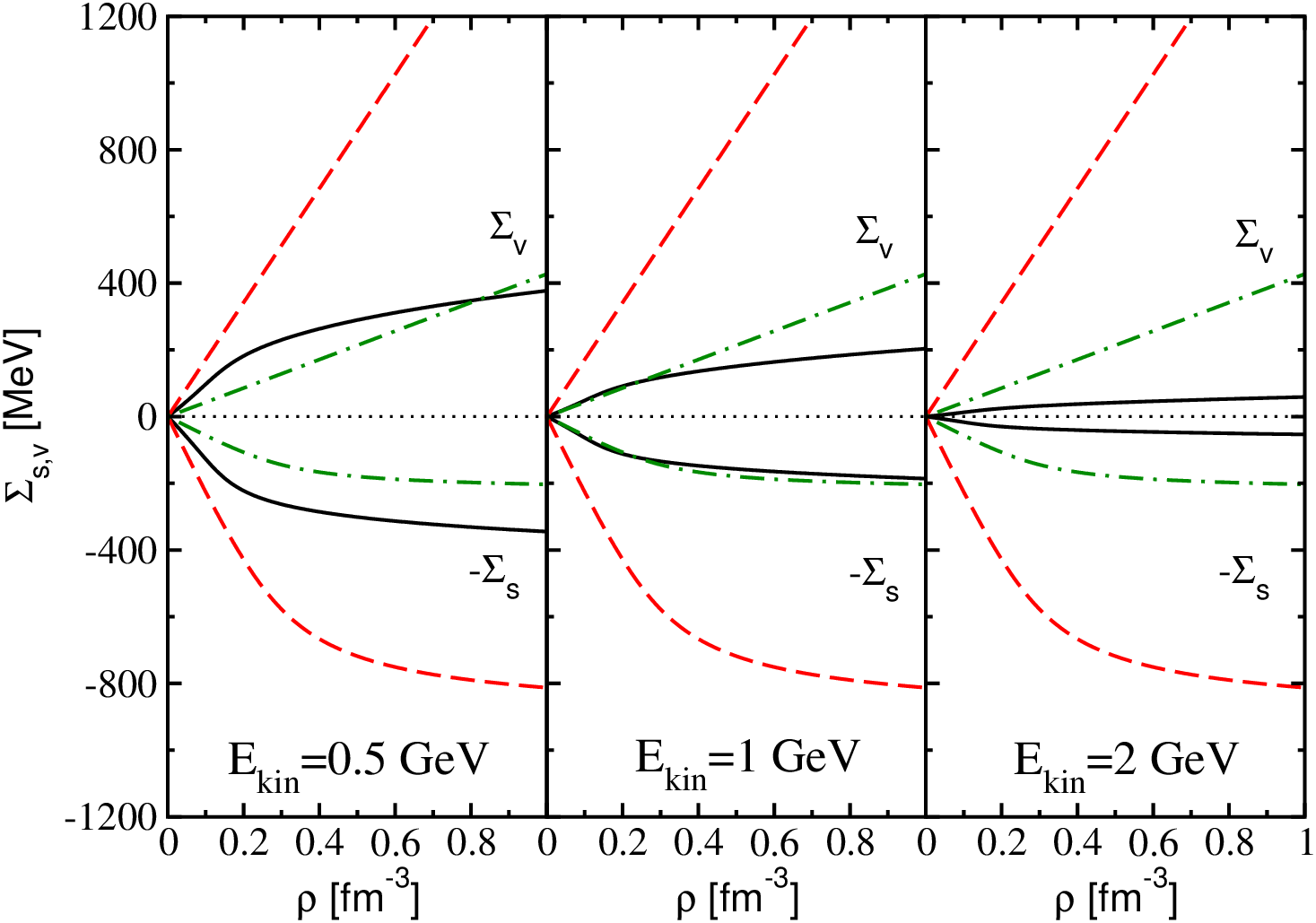}
        \end{center}
        \vspace{-0.2cm}
        \caption{Density dependence of the scalar and vector Lorentz-components 
of the antinucleon self-energy in nuclear matter at energies of 
$E_{\rm kin}= 0.5$ GeV (left), $E_{\rm kin}= 1$ GeV (middle) and
$E_{\rm kin}= 2$ GeV (right)
using the linear Walecka model (dashed lines), the linear Walecka model with 
rescaled couplings with the factor $\xi=0.25$~\cite{larionov} (dash-dotted) 
and the NLD approach (solid lines).}
        \label{fig1a}
        \end{figure*}
For antinucleon interactions in nuclear matter the mean-field potential consists of the sum 
of scalar and vector self-energies. At vanishing momentum and at saturation density 
the linear Walecka model leads to a value of $-\Sigma_{v}-\Sigma_{s}\approx -700$~MeV, 
which is too deep according to phenomenology~\cite{data1,data2}. This feature has been 
always a critical problem in standard RMF models. Even the inclusion of non-linear 
self-interactions of the $\sigma$ field (and eventually of the $\omega$ field)~\cite{boguta} 
do not improve the result, since non-linear self-interactions become pronounced
only above the saturation density. On the other hand, the NLD model reduces 
considerably the deepness of the potential at zero momentum by almost a factor of two.  
The particular difference of the potential depth at vanishing momentum between conventional 
RMF and NLD is not a trivial issue. The consequences of such an energy and density behavior 
will be discussed below when considering the optical potentials.

As discussed in Ref.~\cite{larionov}, in order to reproduce the data from antiproton-induced 
reactions, the antinucleon-meson coupling constants of the Walecka model have to be rescaled 
by a factor of $\xi\simeq 0.2-0.3$. 
Fig.~\ref{fig1} shows also the calculations in the linear Walecka model, but with rescaled couplings 
by a factor of $\xi=0.25$ (dash-dotted curves in Fig.~\ref{fig1}). Indeed, as one can see 
in Fig.~\ref{fig1}, the rescaled Walecka model~\cite{larionov}  reproduces the NLD results 
in average. However, former results fail to reproduce the energy dependence and, in particular, 
the density dependence, as it is demonstrated in Fig.~\ref{fig1a}. In Fig.~\ref{fig1a} 
the density dependence (at various fixed kinetic energies) of the antinucleon 
self-energies is displayed. The NLD self-energies saturate with density and energy 
according to microscopic Dirac-Brueckner studies, as discussed in detail in~\cite{nld}. 
In the conventional Walecka model the vector self-energy diverges with increasing density leading 
to a too strong repulsion at high densities. In fact, this effect of repulsive nature is 
softened in the rescaled model to large extend, however, the linear divergent behavior of the vector 
self-energy still remains. The NLD calculations agree (in average) with the rescaled Walecka model 
around the saturation density and at kinetic energies around $1$~GeV only. 

The very different energy behavior of the self-energies between NLD and linear Walecka models influences 
the Schr\"odinger-equivalent optical potential. In general, 
it is extracted from (anti)proton-nucleus scattering and therefore it is suited for comparisons between 
theory and empirical studies. Its real part is given by
\begin{equation}
 \mathfrak{Re} U_{\rm opt} = \pm\frac{E}{m} \Sigma_{v} - \Sigma_{s}
 + \frac{1}{2m} \left( \Sigma^{2}_{s} - \Sigma_{v}^{2}\right)
\:, \label{U_opt}
\end{equation}
where $E$ is the energy of an (anti)nucleon with bare mass $m$ inside nuclear matter at a fixed 
baryon density and upper (lower) sign holds for nucleons (antinucleons). At first 
we consider the proton-nucleus optical potential. 

        \begin{figure}[t]
        \begin{center}
        \includegraphics[width=0.7\linewidth]{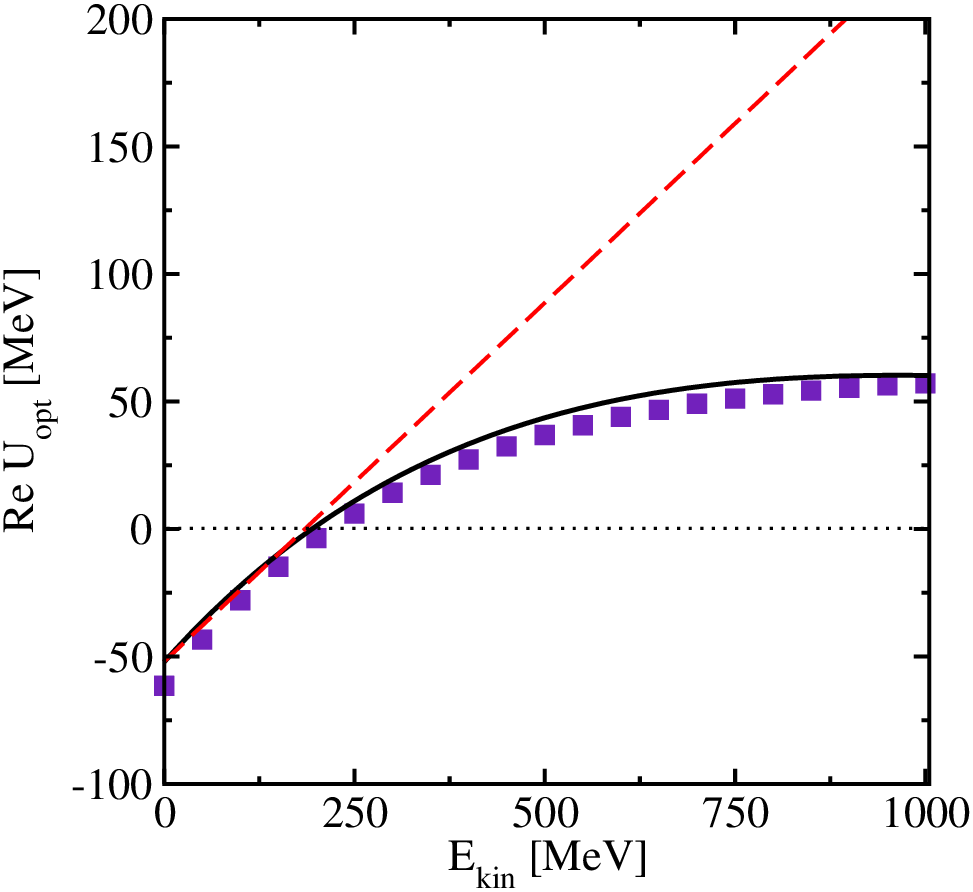}
        \end{center}
        \vspace{-0.2cm}
        \caption{Energy dependence of the Schr\"{o}dinger equivalent 
        proton optical potential at saturation density 
        $\rho_{sat}=0.16~fm^{-3}$. Theoretical calculations in the 
        linear Walecka model (dashed) and NLD approach (solid) are 
        compared to Dirac phenomenology \protect\cite{hama}. }
        \label{fig2}
        \end{figure}

Fig.~\ref{fig2} shows the real part of the optical potential according to Eq.~(\ref{U_opt}) as function of the 
nucleon kinetic energy $E_{kin}=E-m=\sqrt{p^{2}+m^{*2}}+\Sigma_{v}-m$. 
The linear Walecka model (dashed curve) predicts the behavior of the 
optical potential versus energy only qualitatively, and strongly diverges
with increasing kinetic energy of the nucleon. 
It does not reproduce the empirical saturation at higher energies. 
This problem is well known in RMF and has already attracted much attention in the past~\cite{typel}.
 Of course, the main reason for
such strong deviation is the missing energy dependence of the self-energy in the
standard RMF. As 
discussed in detail in Ref.~\cite{nld}, the NLD approach resolves this issue 
of RMF models. The solid curve in Fig.~\ref{fig2} corresponds to the
NLD calculations and describes the data very well.

\begin{figure}[t]
        \begin{center}
        \includegraphics[width=0.7\linewidth]{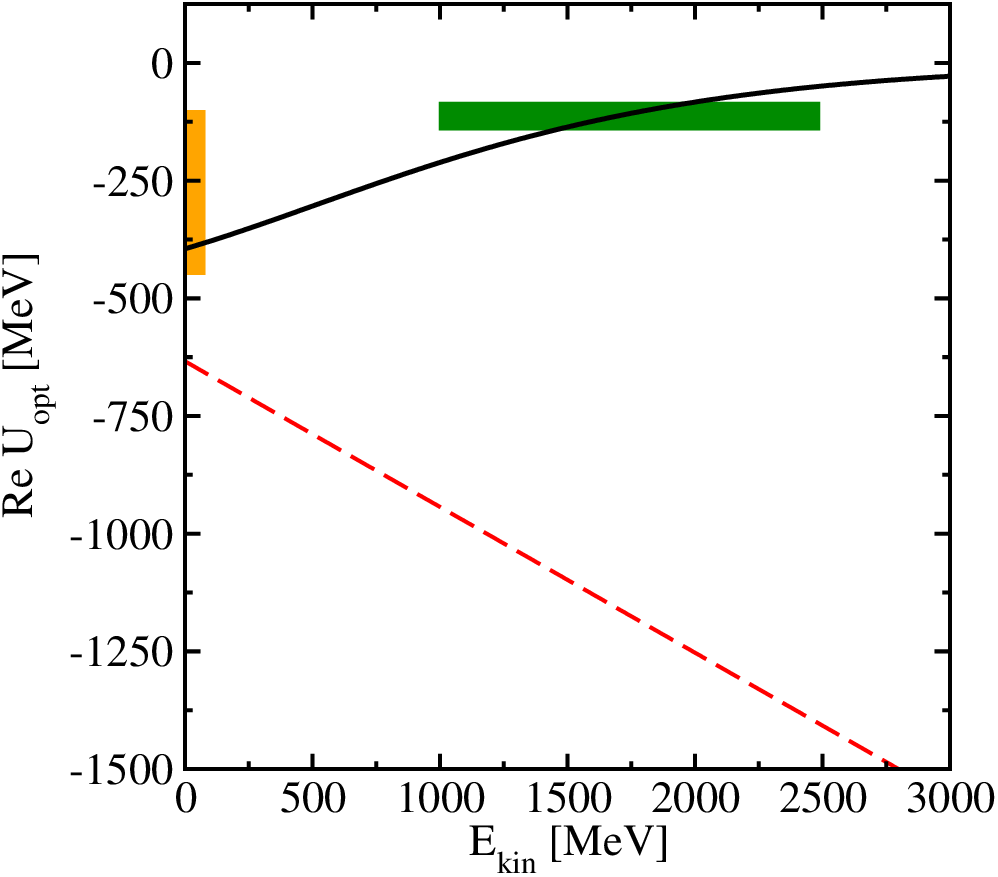}
        \end{center}
        \vspace{-0.2cm}
        \caption{Same as Fig.~\protect\ref{fig2}, but for the antiproton case. The 
        filled box at vanishing momentum represents empirical data extrapolated to saturation 
        density \protect\cite{data1}. The second filled area at kinetic energies between $1000$ 
        and $2500$~MeV is taken from transport calculations on antiproton-nucleus reactions 
        \protect\cite{larionov}.}
        \label{fig3}
        \end{figure}

On the other hand, the interaction of an antinucleon at a given momentum relative to nuclear matter 
at rest is quite different with respect to the proton-nucleus interaction: the sign of the 
Lorentz-vector self-energy changes in Eq.~(\ref{U_opt}). Therefore, in the linear Walecka model 
the real part of the optical potential is again a linear function in energy,
as in the nucleon case, but 
now it diverges to $-\infty$ (see Fig.~\ref{fig3}, dashed curve). 
Such a prediction is in contradiction with calculations using dispersion relations \cite{cass2}. 
In fact, by fitting the imaginary part of the antinucleon-nucleus optical potential to the total 
proton-antiproton cross section, its real part decreases with increasing energy. 
Furthermore, an existing information from heavy-ion collisions~\cite{cass2} and 
reactions induced by protons~\cite{cass2} and antiprotons~\cite{larionov} give
clear evidence for a 
considerable reduction of the antiproton-nucleus optical potential with rising energy. 
As has been discussed in Ref.~\cite{larionov}, the transport theoretical description of 
antiproton-nucleus data is  not possible within the conventional Walecka
model, except if one rescales the 
antinucleon-meson coupling constants by a phenomenological factor of $\xi\approx 0.2$. This
is not compatible with G-parity arguments and suggests a strong violation of
the charge conjugation symmetry in the nuclear medium~\cite{mishustin}, which
otherwise must be a perfect symmetry in strong interactions. 

On the contrary, 
the NLD calculations (solid curve in Fig.~\ref{fig3}) predict a completely
different behavior as compared with the Walecka model. It results in a much 
softer potential at vanishing momentum and much stronger decrease of the real part
of the optical potential $\mathfrak{Re}U_{\rm opt}$ with increasing energy. 

Due to the large annihilation cross section experimental data at 
low energies can be obtained only at very low densities 
$\rho\simeq (0.005\div 0.02) \rho_{\rm sat}$ 
close to the nuclear surface~\cite{data1,data2}, while empirical information at saturation 
density is obtained by extrapolation only. 
At these low densities the NLD model leads to values of $\mathfrak{Re}
U_{opt}\simeq -(6\div 50)$~MeV, 
which seem to be still too deep with respect to the data~\cite{data1,data2}. 
At the density of interest $\rho=\rho_{\rm sat}$ the NLD model predicts a
rather soft potential, which is much closer to extrapolated data~\cite{data1,data2} and dispersion 
relations~\cite{cass2} (filled box at zero kinetic energy in Fig.~\ref{fig3}). A comparison 
between our model and phenomenological antiproton-nucleus reactions at higher energies seems 
more meaningful. In fact,  with increasing energy the annihilation cross section 
drops strongly and it is supposed that the antiprotons may penetrate deeper 
inside the nuclear interior, and thus densities 
close to $\rho_{\rm sat}$ can be  tested. The second filled area in Fig.~\ref{fig3} 
shows the empirical optical potential as extracted from the transport theoretical analyses 
in Ref.~\cite{larionov,larionov2}. In this energy region the comparison
between NLD results and transport calculations (which use conventional RMF, but
with largely reduced antinucleon-meson couplings) turns out to be fairly
well. Our results are also in qualitative agreement with 
the analysis of Ref.~\cite{Zhang}, where a strong decrease of 
$\mathfrak{Re}U_{\rm opt}$ with increasing energy is obtained. 

Interestingly, the antinucleon optical potentials $\mathfrak{Re}U_{\rm opt}$  strongly differ at zero momentum
between NLD and standard RMF, while in the nucleon case (see Fig.~\ref{fig2}) 
no differences were visible. By considering the fields at the same baryon density and at 
zero momentum one would naively expect a similar potential depth for both models. Indeed, 
the non-linear effects start to dominate above the saturation density~\cite{nld}. However, 
the observed difference at zero momentum comes from the in-medium dependence of the (anti)nucleon 
single-particle energy. At fixed saturation density the 
energy shift, caused by the difference (proton-nucleus) or sum (antiproton-nucleus) of two big 
fields, varies strongly between the two models. However, small shift variations in the energy 
affect the NLD self-energies, 
due to their pronounced energy dependence. On the other hand, the fields of the linear Walecka 
model are not influenced due to their independence on energy. This feature becomes more pronounced 
with increasing density, as seen in Fig.~\ref{fig1} (middle and right panels). In terms of the 
optical potentials the interpretation is similar. In the proton-nucleus case (Fig.~\ref{fig2}) 
the slopes between both models at vanishing momentum are essentially the
same. Therefore the in-medium energy shift 
is of minor relevance and there is no gap between both potentials. In the 
antiproton-nucleus case (Fig.~\ref{fig3}) the gap is much more pronounced due to the quite different 
slopes between NLD and linear Walecka optical potentials. 

        \begin{figure}[t]
        \begin{center}
        \includegraphics[width=0.7\linewidth]{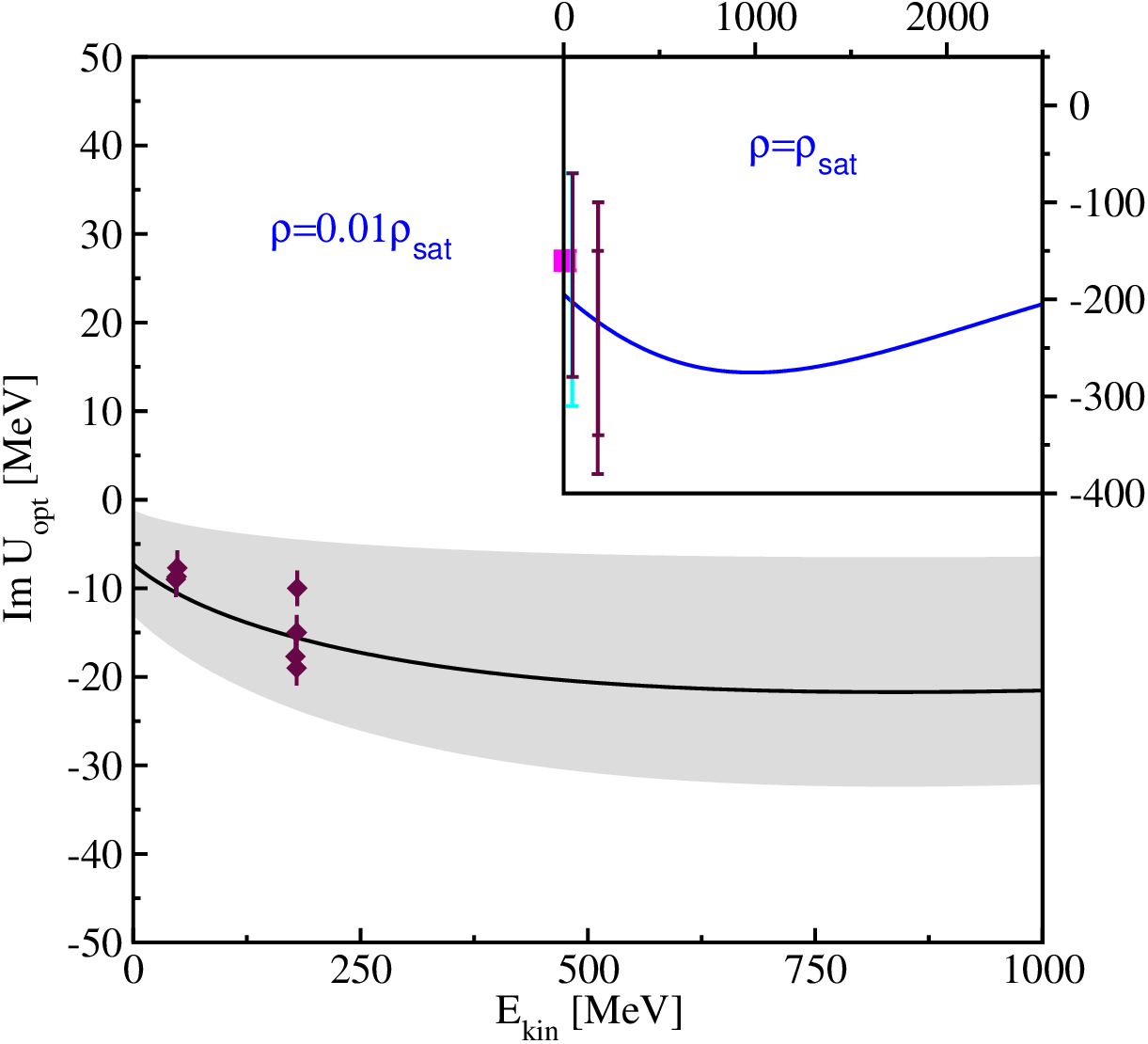}
        \end{center}
        \vspace{-0.2cm}
        \caption{Energy dependence of the imaginary part of the antinucleon 
		optical potential at low density, as indicated. The theoretical result, extracted 
		from the dispersion relation (see Eq.~(\protect\ref{DispRel})) 
		in the NLD approach (solid curves) is compared to experimental data (symbols) taken from 
		\protect\cite{data2}. The filled area indicates the model changes by varying the density from 
		$0.005$ up to $0.02$ (relative to $\rho_{sat}$). 
		The inserted panel shows the same quantity but at the fixed saturation 
		density $\rho_{sat}=0.16~fm^{-3}$, again in comparison to data (symbols), which are 
		extrapolated to $\rho_{sat}$.}
		\label{fig4}
        \end{figure}

For a complete description of in-medium antiproton interactions also the imaginary part of the 
optical potential is needed. Since RMF does not provide the imaginary part of the self-energies, 
we estimate $\mathfrak{Im}U_{\rm opt}$ using dispersion relation~\cite{Disp}
\begin{equation}
\mathfrak{Im}U_{\rm opt}(p) = -\frac{2p}{\pi} \; {\cal P} \int_{0}^{\infty} 
 \frac{\mathfrak{Re}U_{\rm opt}(p^{\prime \;})}{p^{\prime\;2}-p^{2}} dp^{\prime}
\:, \label{DispRel}
\end{equation}
where $p\equiv |\vec{p}\;|$ means the antiparticle momentum and ${\cal P}$ denotes the principal value. 
The real part $\mathfrak{Re}U_{\rm opt}$ 
is taken from the NLD model. The results are shown in Fig.~\ref{fig4}. The insert in Fig.~\ref{fig4} 
shows $\mathfrak{Im}U_{\rm opt}$ versus the kinetic in-medium energy $E_{kin}$ at saturation density. At vanishing 
kinetic energy the imaginary part of the optical potential is rather large
($\simeq -200$~MeV) and consistent with empirical 
information (error bars)~\cite{data2}. 
At high beam energies $\mathfrak{Im}U_{\rm opt}$ start to decrease again, but 
remains rather strong. The present estimation seems to be in line with the empirical study 
of Ref.~\cite{Zhang}, where $\mathfrak{Im}U_{\rm opt}=-135$~MeV is essentially independent on the energy 
in the range from $180$~MeV up to $\simeq 2$~GeV. 

According to Refs.~\cite{data1,data2} antiprotons 
penetrate the nuclear surface up to densities of $\rho\simeq (0.005-0.02)\rho_{\rm sat}$ before annihilation. 
Therefore, we calculate $\mathfrak{Im}U_{\rm opt}$ at this low densities, as shown in the main panel of 
Fig.~\ref{fig4}. The filled area indicates the model calculations for matter
densities $\rho\simeq (0.005-0.02)\rho_{sat}$. The solid curve shows the model
result at an average density of 
$\rho \simeq 0.01~\rho_{sat}$. As one can see the NLD model reproduces the data~\cite{data2}
fairly well also at these low densities. 

\section{Summary and Outlook}
In summary, the NLD model, which incorporates on a mean-field level non-linear 
effects in baryon density and simultaneously in single-particle energy, has been applied 
to nucleon and antinucleon interactions in nuclear matter. 
We have shown that due to the explicit energy dependence of the self-energies the proton-nucleus 
optical potential is very well reproduced. 
At the same time, the NLD model predicts a much softer
real part of the antiproton optical potential at low energies as compared to
the Walecka model. We also find a strong decrease of the optical potential
with increasing energy. These results are remarkably
consistent with available information from reactions involving heavy ion and
(anti)proton beams and other studies based on dispersion-theoretical
approaches. A comparison with the conventional Walecka model has shown that the
main effect responsible for a description of the in-medium
(anti)nucleon optical potential originates from the energy dependence of the
mean-field, which is absent in standard RMF models. We further estimated 
the imaginary part of the antiproton optical potential 
within the NLD model using dispersion relation. The results 
were in qualitative agreement with the low density data and empirical extrapolations at saturation density. 
We, therefore, conclude that RMF models may remain a very useful theoretical tool for the
description and analysis of the antinucleon interactions in nuclear medium.  

The energy dependence of the real and imaginary parts of the antinucleon optical potential, 
studied in this work, is expected to be important at energies relevant 
for the \panda experiment at FAIR.  The nuclear compression due to the strong 
attractive antinucleon mean-field, which significantly differs between the Walecka and the NLD 
models, and also the very different energy behavior between them will affect the dynamics 
of antiproton-nucleus reactions. Thus, we expect various observable phenomena at FAIR as 
important probes for the NLD predictions. For instance, the fragmentation of the excited and radially 
expanded residual nuclei, where the energy transferred to radial expansion is expected to depend 
on the degree of compression. Strangeness production of $s=-1$ and especially $s=-2$ hyperons, 
such as cascade ($\Xi$) particles, is expected to be medium dependent, in particular close to 
threshold energies. The associate formation of single-$\Lambda$ and, in particular, of 
double-strange hypernuclei is thus supposed to be also model dependent. 
As an outlook we conclude the importance and relevance of our results for the future activities at FAIR.

\section*{Acknowledgements}
This work was supported by HIC for FAIR, DFG through TR16 and by BMBF. 
We acknowledge useful discussions with A.~Larionov.




\begin{thebibliography}{99}
\biboptions{sort&compress}

\bibitem{EoS}
  N.~Herrmann, J.P.~Wessels, T.~Wienold, 
  Annu.\ Rev.\ Nucl.\ Part.\ Sci.\  49 (1999) 581.

\bibitem{pheno}
	B.~Bl\"attel, V.~Koch, U.~Mosel, 
	Rep.\ Prog.\ Phys.\ 56 (1003) 1.

\bibitem{micro}
  	B.~ter~Haar, R.~Malfliet, 
    Phys.\ Rep.\ 149 (1987) 207.

\bibitem{cass1}
  K.~Weber \textit{et al.}, 
  Nucl.\ Phys.\ A539 (1992) 713.

\bibitem{lear}
C.~Amsler, Annu.\ Rev.\ Nucl.\ Part.\ Sci.\  38 (1991) 219.

\bibitem{cass2a}
W.~Cassing, E.L.~Bratkovskaya, 
Phys.\ Rep.\ 308 (1999) 65.

\bibitem{cass2}
	S.~Teis \textit{et al.}, 
	Phys.\ Rev.\ C50 (1994) 388.
	
\bibitem{oset}
  E.~Hernandez, E.~Oset,
  Nucl.\ Phys.\  A493 (1989)  453;
  E.~Hernandez, E.~Oset,
  Nucl.\ Phys.\  A455 (1986)  584-601.
	
\bibitem{gal}
C.J.~Batty, E.~Friedman, A.~Gal, 
Phys.\ Rep.\ 287 (1997) 385.

\bibitem{panda_big}
\panda collaboration, hep-ex/0903.3905v1.

\bibitem{panda}
T.~Johannsson (\panda collaboration), 
XLVIII International Winter Meeting on Nuclear Physics, BORMIO2010
January 25-29, 2010, Bormio, Italy.

\bibitem{abhf}
T.~Suzuki, H.~Narumi, Nucl.\ Phys.\ A426 (1984) 413.

\bibitem{wal74}
H.P.~Duerr, Phys.\ Rev.\ 103 (1956) 469;\\
J.D. Walecka, Ann.\ Phys.\ (N.Y.) 83 (1974) 497.

\bibitem{heidenbauer}
T.~Hippchen et al., Phys.\ Rev.\ C44 (1991) 1323.

\bibitem{larionov}
A.B.~Larionov \textit{et al.}, Phys. Rev. C80 (2009) 021601(R).

\bibitem{mishustin}
I.N.~Mishustin \textit{et al.}, Phys.\ Rev.\ C71 (2005) 035201.

\bibitem{nld}
T.~Gaitanos, M.~Kaskulov, U.~Mosel, 
Nucl.\ Phys.\ A828 (2009) 9.

\bibitem{dbhf}
C.J.~Horowitz, B.D.~Serot, 
Nucl.\ Phys.\ A464 (1987) 613.

\bibitem{data1}
P.D.~Barnes \textit{et al.}, Phys.\ Rev.\ Lett.\ 29 (1972) 1132;\\
H.~Poth \textit{et al.}, Nucl.\ Phys.\ A294 (1978) 435;\\
C.J.~Batty, Nucl.\ Phys.\ A372 (1981) 433;\\
C.-Y.~Wong \textit{et al.}, Phys.\ Rev.\ C29 (1984) 574.

\bibitem{data2}
E.~Friedmann, A.~Gal, J.~Mares, Nucl.\ Phys.\ A761 (2005) 283;\\
S.~Janouin {\it et al.}, Nucl.\ Phys.\ A451 (1986) 541.

\bibitem{boguta}
J.~Boguta, A.R.~Bodmer, Nucl.\ Phys.\ A292 (1977) 413;\\
Y.~Sugahara and H.~Toki, Nucl.\ Phys.\ A579 (1994) 557.

\bibitem{hama}
E.D.~Cooper \textit{et al.}, 
Phys.\ Rev.\ C47 (1993) 297.

\bibitem{typel}
T.~Maruyama et al., Nucl.\ Phys.\ A573 (1994) 653;\\
P.K.~Sahu et al., Nucl.\ Phys.\ A672 (2000) 376;\\
S.~Typel, Phys.\ Rev.\ C71 (2005) 064301.

\bibitem{larionov2}
A.B.~Larionov, private communication.


\bibitem{Zhang}
Z.~Yu-shun \textit{et al.}, Phys.\ Rev.\ C54 (1996) 332.

\bibitem{Disp}
J.~Rammer, \textit{Quantum Field Theory of Non-Equilibrium States}, 
Cambridge University Press (July 30, 2007).

\end{thebibliography}
\end{document}